# Two-Factor Biometric Verification with ECG: Two Cancelable Approaches

Jui-Kun Chiu*, Tzu-Yun Lin, Wei-Shen Hsu and Shun-Chi Wu *Member, IEEE*

*Abstract* —Biometric authentication relies on an individual's physiological or behavioral traits to verify their identity before granting access permission to a system or device without remembering anything. Although electrocardiograms (ECGs) have been considered a biometric trait, an ECG biometric recognition system that operates in verification mode is rarely considered. This study proposes two two-factor cancelable biometric verification schemes that enable identity recognition using ECGs. Using bioconvolving and minimum average correlation energy biometric filters, revocable and irreversible templates can be constructed to avoid privacy invasion and security concerns associated with ECG biometric recognition. An interquartile range-based method is adopted to determine if an identity match exists, enabling identity verification under the influence of inter-beat variation. The experimental results indicated that the proposed schemes could achieve an equal error rate as low as 1% when the inter-beat variation was properly addressed.

## I. INTRODUCTION

Authentication is the process of identifying users who request access to a system, network, or device, and access permission is granted through identity recognition based on passwords. Constructing a strong password requires sufficient letters, numbers, and symbols in a sequence to make it challenging to crack [1]. However, this complexity may make it difficult to remember passwords. Biometric authentication relies on the unique physiological or behavioral traits of individuals to verify their identities. Unlike passwords, biometric traits are associated with individuals who do not remember anything. A single trait or multiple traits can be used for biometric authentication systems, depending on the desired infrastructure and level of security. Various biometric traits currently in use include fingerprints, faces, and irises. Although these traits are widely used in many applications, their extrinsic nature makes them susceptible to falsification. For instance, a contact lens can spoof an iris recognition system with a printed iris image [2], [3]. Fake facial masks can attain significant attack performance on facial recognition systems with advancements in three-dimensional reconstruction and printing technologies [4], [5]. Therefore, the search for new biometric attributes is ongoing, and electrocardiograms (ECGs), the recordings of the electrical activity of the heart, have gained interest [6-8].

ECG has several required characteristics, such as universality, distinctiveness, permanence, and collectability,

This work was supported by the Ministry of Science and Technology of Taiwan, R.O.C., under the contracts MOST 109-2221-E-007-012-MY3.

J.-K. Chiu (*corresponding author, chiu.steve98@gmail.com), T.-Y. Lin (evening.primrose321@ gmail.com), W.-S. Hsu (w05200520@gmail.com), and S.-C. Wu (shunchi.wu@mx.nthu.edu.tw) are with the Department of Engineering and System Science, National Tsing Hua University, Hsinchu, Taiwan 30013.

to be a biometric attribute. Moreover, its intrinsic and dynamic nature makes it difficult to steal and forge. Various approaches that utilize ECG for biometric recognition include fiducial feature- [9], principal component (PC)- [10], and wavelet transform-based [11] approaches. A common characteristic of these approaches is that ECGs must be transformed into abstract representations referred to as biometric templates before being applied to recognition tasks. However, the application of the approaches is accompanied by security and privacy concerns. For instance, a very high correlation coefficient can be obtained by evaluating the similarity between the original ECG waveforms and those reconstructed from the PCs of a selected subject [12], leading to privacy invasion. Furthermore, the fiducial features of a subject can be used to synthesize ECGs that can successfully attack a recognition system [13].

An approach for addressing privacy and security concerns involves building an ECG recognition scheme based on deep learning [14]. Because the discriminant information between different subjects' ECGs is learned and embedded within the multiple hidden layers of the network model, the need for template construction and the risk of a template being stolen are eliminated. Another approach is to construct cancelable templates for which recovery is infeasible due to blending artificial patterns into the given biometric samples [7]. Because reconstruction is infeasible, the private information of a stolen template can be protected, and reconstructing ECGs for system attacks is not possible. Moreover, because the template is generated by blending a random pattern with a given ECG sample, it can be revocable like passwords once stolen. This study aims to develop cancelable ECG biometric recognition approaches that operate in verification mode. Before verification, a subject must claim their identity, and the system validates the identity by comparing the provided ECGs with the corresponding template stored in the system. Based on the concepts of bioconvolving [15] and the minimum average correlation energy (MACE) biometric filter [16], revocable and irreversible templates are constructed to avoid privacy invasion and security concerns.

## II. PROPOSED ECG BIOMETRIC VERIFICATION SCHEMES

### A. Data Preprocessing

Baseline wanders and power-line interference are artifacts making ECG interpretation difficult. Extracting baseline wander can be achieved via two median filters with window sizes of 200 *ms* and 600 *ms*, respectively. A second-order IIR notch filter can be used to remove the power-line interference. After artifact removal, the noise-free ECGs are cut into segments of $T_d$ *s* centered at the R peaks. Depending on the sampling rate $f_s$, each beat segment has a length $l = T_d f_s$.

## B. Cancelable Template Construction

Here, we present two cancelable template construction schemes. We assume that $N_{en}$ beat segments are collected during enrollment for each subject.

### 1) Bioconvolving

To construct a template for a subject to be enrolled through bioconvolution, we first cut the ensemble average of the $N_{en}$ beat segments, $\bar{\mathbf{x}} \in \mathbb{R}^{l \times 1}$, into $k$ pieces of equal length, $l/k$. Thereafter, a random key, $\mathbf{r}_{bc} = [r_1, r_2, \cdots, r_k]$, is generated, where $r_i$ is a random integer drawn from the range $[1, l/k–1]$, which is used to further divide the $i^{th}$ ECG piece into two parts of lengths $r_i$ and $l/k–r_i$. We compute the convolution of these two parts and repeat the process of cutting and convolution for $k$ ECG pieces. Thereafter, template $\mathbf{F}$ is constructed by stacking the $k$ convolution results as a vector of size $(l–k) \times 1$.

### 2) MACE biometric filter

First, we convolve the $N_{en}$ beat segments of a subject to be enrolled with a random key, $\mathbf{r}_{mace} \in \mathbb{R}^{k \times 1}$ and apply the discrete Fourier transform to the convolution results to obtain the encrypted ECG segments, $\mathbf{m}_i \in \mathbb{C}^{(l+k-1) \times 1}$ for $i = 1, \ldots, N_{en}$. To construct the MACE biometric filter $\mathbf{h} \in \mathbb{C}^{(l+k-1) \times 1}$ for the subject, we use [17]

$$\mathbf{h} = \mathbf{D}^{-1}\mathbf{M}(\mathbf{M}^H \mathbf{D}^{-1}\mathbf{M})^{-1}\mathbf{u}, \quad (1)$$

where $\mathbf{M} \in \mathbb{C}^{(l+k-1) \times N_{en}}$ is the stack of the $N_{en}$ encrypted ECG segments; $\mathbf{D}$ is a diagonal matrix of size $l+k–1 \times l+k–1$ whose diagonal elements are the magnitude square of the associated element of $\bar{\mathbf{m}}$ and the ensemble average of $\mathbf{m}_i$; the superscript $H$ denotes the conjugate transpose of a matrix; $\mathbf{u} \in \mathbb{R}^{(l+k-1) \times 1}$ is a user-specified constraint vector whose elements are 1 in this study. A crucial characteristic is that the cross-correlation of the MACE biometric filter, $\mathbf{h}$, and the given encrypted ECG segment, $\mathbf{m}$, exhibits a sharp peak at the origin of the correlation spectrum if $\mathbf{h}$ and $\mathbf{m}$ are generated from the beat segments of the same subject, as shown in the left Figure 1(a). If not, the spectrum will have more than one peak with a significantly low correlation, as shown in the right of Figure 1(a).

The progression of the proposed scheme is illustrated in Figure 1(b). Template $\mathbf{F}$, MACE filter $\mathbf{h}$, and the correlation spectra of $\mathbf{h}$ and $\mathbf{m}$ must be stored in the system database for identity verification. The revocability of $\mathbf{F}$ and $\mathbf{h}$ is ensured because they can be replaced by new constructed by applying different random keys. Irreversibility is also possessed by $\mathbf{F}$ and $\mathbf{h}$ because recovering the beat segment from $\mathbf{F}$ or $\mathbf{h}$ is infeasible without a random key. Notably, the random key is possessed only by the subject to be enrolled in the proposed system. Finally, either $\mathbf{F}$ or $\mathbf{h}$ is generated by convolving the segmented beat pieces or beat segment with a random key. This makes a beat segment "smeared" such that the physiological and pathological conditions of the subject cannot be disclosed from $\mathbf{F}$ and $\mathbf{h}$.

## C. Identity Verification for an Enrollee

When an enrolled subject tries to access the system, they have to claim their identity, present the random key, and provide $N_v$ beat segments for identity verification. The beat segments and random key are used to construct a bioconvolving template or compute a mean-encrypted ECG segment following the procedure described in Section II.B. The stored template $\mathbf{F}$ or MACE biometric filter $\mathbf{h}$ of the subject is retrieved from the database according to the claimed identity to "compare" with the newly formed template or the mean encrypted ECG segment. If the "similarity" between them is lower than the preset threshold of the claimed subject, the subject requesting access to the system is deemed to be who they claim; otherwise, the subject is considered to be unregistered. The flow diagram in Figure 1(b) illustrates the proposed verification scheme. All databases are established during enrollment.

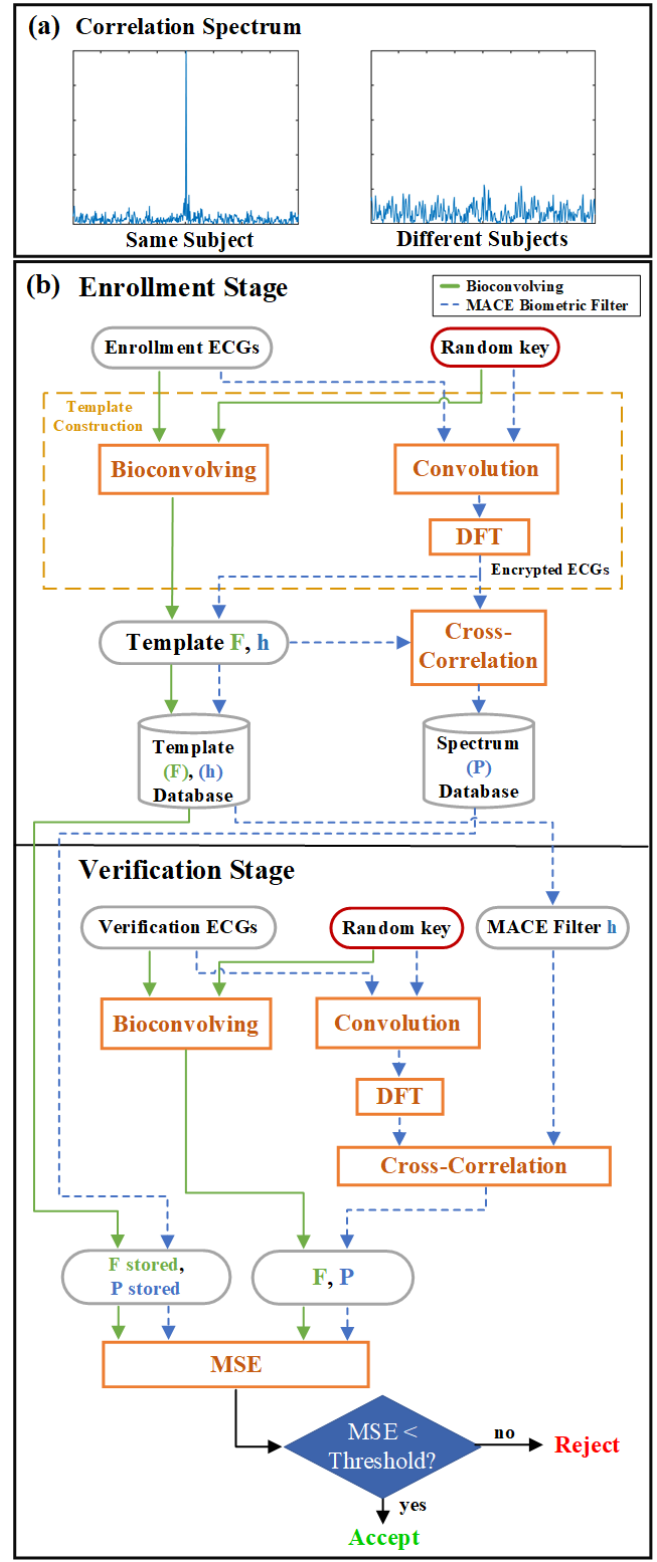

Fig. 1. Algorithm flowchart of the proposed scheme for unknown subject identity verification using ECGs.

We use the mean squared error (MSE) as a measure of similarity. However, the calculation is different for systems based on bioconvolving and the MACE biometric filter. For the former, MSE is the error between the newly formed template **F** and that retrieved from the system. For the latter, the newly formed mean-encrypted ECG segment is first used to evaluate the cross-correlation with the MACE filter **h** belonging to the claimed subject. Thereafter, the resulting correlation spectrum is used to calculate the MSE of the data retrieved from the system. Although the MSE is ideally zero if the subject is genuinely who they claim, variations in MSE from one access trial to another are difficult to avoid owing to inter-beat variability. To determine if a match exists, an interquartile range (IQR)-based method is adopted. IQR is the difference between the first and third quartiles of a set of obtained MSEs, denoted as $Q_1$ and $Q_3$, respectively. Notably, we have $N_{en}$ beat segments that can be used for MSE evaluation for any subject enrolled. The threshold is then set based on $Q_1$ and $Q_3$ of the obtained MSEs

$$T_{mse} = Q_3 + k_{iqr} \times (Q_3 - Q_1), \quad (2)$$

where $k_{iqr}$ is the parameter that controls the size of the fence region. Each enrolled subject has their own threshold. Later, if the MSE obtained for a subject to access the system is lower than $T_{mse}$, the subject is deemed who they claim; otherwise, the subject is considered unregistered.

### III. EXPERIMENTS AND RESULTS

#### A. Dataset

The proposed schemes' performance was evaluated using the Lead-I recordings of 285 subjects from the Pysikalisch-Technische Bundesanstalt (PTB) database [18]. Each ECG record was sampled at 1 kHz with a 12-bit resolution. Before proceeding, each ECG record was subjected to 50 Hz power-line interference suppression and baseline wander removal. The noise-free ECGs were then segmented with respect to the detected R peaks with a length of 0.8 s. Finally, the obtained beat segments were randomly selected for enrollment or verification in the following experiments.

#### B. Results and Discussion

A verification system should verify the identity of a registered subject and exclude the unregistered subjects. Two types of errors are possible during verification: (1) an unregistered subject is incorrectly accepted and (2) a registered subject is incorrectly rejected. The false-positive rate (FPR) and false-negative rate (FNR) are the metrics for these events. To evaluate the FPR, we adopted a leave-one-out strategy in which one of the 285 subjects was selected as the registered subject and the remaining subjects intruded their account. We repeated this process ten times for each subject, and the FPR was calculated as the ratio of the number of false accepted trials to the total number of trials. The FNR was calculated as the ratio of the number of incorrect rejections to the number of total trials, where each registered subject accessed their account ten times. Notably, FPR and FNR vary with $k_{iqr}$, and a $k_{iqr}$ corresponding to a low FNR may incur a high FPR. To determine a suitable $k_{iqr}$, the usual choice is a value that yields an equal error rate (EER) (*i.e.*, FPR = FNR). The lower the EER, the higher the accuracy of the biometric system is.

Two factors, the sampling rate and number of beats used for verification, that could affect the verification accuracy were investigated. Different sampling rates were achieved using a low-pass anti-aliasing filter on the original recordings followed by decimation. The low-pass filter was a Chebyshev Type-I filter of order eight with a normalized cutoff frequency of $0.8\pi/r$ and a passband ripple of 0.05 dB, where r is the downsampling factor. Typically, the constraint on the time required for enrollment is not necessarily strict, and additional heartbeats from the enrollee can be acquired. However, this is not true during verification. Since acquiring one complete heartbeat requires approximately 1 s of data, collecting more heartbeats unavoidably hinders the identification progress. Thus, the number of beats for identity verification, $N_v$ is set to be smaller than that used for enrollment, $N_{en}$. Several $N_{en}/N_v$ combinations were evaluated: 1) $N_{en}$ = 15 and $N_v$ = 1; 2) $N_{en}$ = 15 and $N_v$ = 3; 3) $N_{en}$ = 15 and $N_v$ = 5; 4) $N_{en}$ = 15 and $N_v$ = 10.

The EERs of the bioconvolving and MACE filter-based schemes are shown in Figures 2 and 3. Unlike fingerprints, ECGs are dynamic signals that may suffer from inter-beat variation owing to changes in the physical state of the subject, which can degrade the verification accuracy. This was evident when $N_v$ was one, as shown in Figures 2 and 3. The effect of inter-beat variation can be mitigated through beat averaging [12]. When $N_v$ was greater than one, the EERs were instantaneously reduced to less than 1% in both schemes. First, one may think that reducing the sampling rate could lead to the loss of morphological details of the ECG waveforms and cause a decline in discrimination power and thus higher EERs. However, as shown in Figures 2 and 3, the EERs decreased as the sampling rate decreased. Notably, in this study, the sampling rate reduction was achieved using a low-pass anti-aliasing filter on the original recordings, followed by decimation. This low-pass anti-aliasing filter provided a means to mitigate the inter-beat variation that made both approaches robust to inter-beat variation and achieved better EERs even when $N_v = 1$. Although the low-pass filter discarded the morphological details of the ECGs of the subject to be enrolled, the loss of discrimination power was compensated by introducing a user-specific random key. The EERs obtained when $N_v$ increased further also verified this point. The EERs were stable regardless of the sample rates and $N_v$ values if the inter-beat variation was properly addressed. Finally, the bioconvolving-based scheme performed better than the scheme based on the MACE biometric filter, particularly when $N_v$ was small. This was attributed to the additional smoothing resulting from the convolution between the pieces of ECGs.

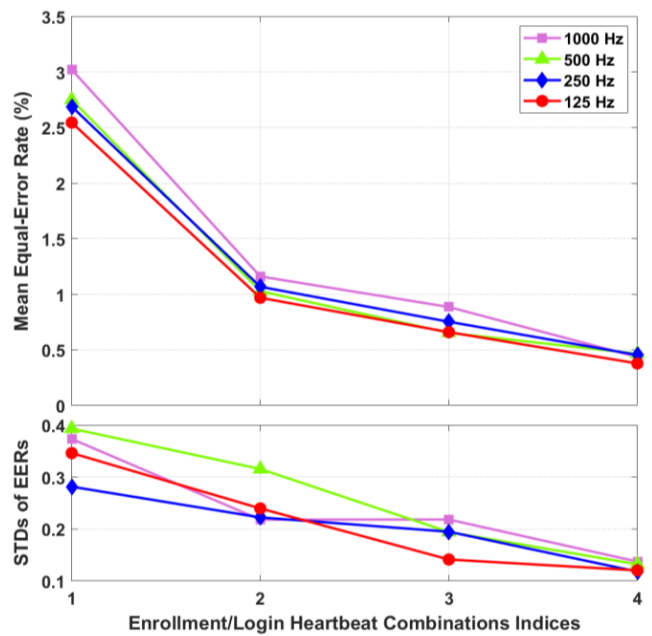

Fig. 2. EERs of the bioconvolving-based approach under different $f_s$ and $N_{en}/N_v$ combinations.

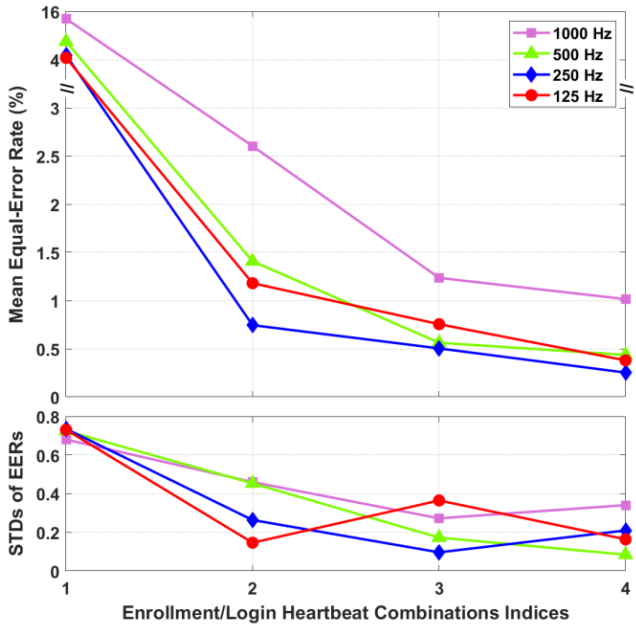

Fig. 3. EERs of the MACE filter -based approach under different $f_s$ and $N_{en}/N_v$ combinations.

## IV. Conclusions

This study presents two two-factor cancelable biometric verification schemes that allow identity recognition using ECG beat segments. The revocability of the bioconvolving templates and MACE biometric filters is guaranteed by applying different random keys when needed. Their irreversibility arises from the fact that recovering the original beat segment from a stolen template or MACE filter is infeasible without a random key for their generation. In addition, convolving a beat segment with a random key makes the distinct components of a beat segment "smeared"; thus, the physiological and pathological conditions of the subject can be avoided. Finally, the IQR-based method is adopted to determine whether an identity match exists. The performance of the proposed schemes was evaluated using ECGs from the PTB database. The results indicated that both schemes achieved an EER as low as 1%.

## References

[1] J. A. Cazier and B. D. Medlin, "Password security: An empirical investigation into e-commerce passwords and their crack times," *Information Systems Security,* vol. 15, no. 6, pp. 45-55, 2006.

[2] A. Morales, J. Fierrez, J. Galbally, and M. Gomez-Barrero, "Introduction to iris presentation attack detection," in *Handbook of Biometric Anti-Spoofing*: Springer, 2019, pp. 135-150.

[3] Z. Wei, X. Qiu, Z. Sun, and T. Tan, "Counterfeit iris detection based on texture analysis," in *2008 19th International Conference on Pattern Recognition*, 2008: IEEE, pp. 1-4.

[4] N. Erdogmus and S. Marcel, "Spoofing 2D face recognition systems with 3D masks," in *2013 International Conference of the BIOSIG Special Interest Group (BIOSIG)*, 2013: IEEE, pp. 1-8.

[5] N. Erdogmus and S. Marcel, "Spoofing face recognition with 3D masks," *IEEE transactions on information forensics and security,* vol. 9, no. 7, pp. 1084-1097, 2014.

[6] M. Ingale, R. Cordeiro, S. Thentu, Y. Park, and N. Karimian, "Ecg biometric authentication: A comparative analysis," *IEEE Access,* vol. 8, pp. 117853-117866, 2020.

[7] S.-C. Wu, P.-L. Hung, and A. L. Swindlehurst, "ECG Biometric Recognition: Unlinkability, Irreversibility, and Security," *IEEE Internet of Things Journal,* vol. 8, no. 1, pp. 487-500, 2020.

[8] S.-C. Wu, P.-T. Chen, A. L. Swindlehurst, and P.-L. Hung, "Cancelable biometric recognition with ECGs: subspace-based approaches," *IEEE Transactions on Information Forensics and Security,* vol. 14, no. 5, pp. 1323-1336, 2018.

[9] Y. Wang, F. Agrafioti, D. Hatzinakos, and K. N. Plataniotis, "Analysis of human electrocardiogram for biometric recognition," *EURASIP journal on Advances in Signal Processing,* vol. 2008, pp. 1-11, 2007.

[10] J. M. Irvine, S. A. Israel, W. T. Scruggs, and W. J. Worek, "eigenPulse: Robust human identification from cardiovascular function," *Pattern Recognition,* vol. 41, no. 11, pp. 3427-3435, 2008.

[11] A. D. Chan, M. M. Hamdy, A. Badre, and V. Badee, "Wavelet distance measure for person identification using electrocardiograms," *IEEE transactions on instrumentation and measurement,* vol. 57, no. 2, pp. 248-253, 2008.

[12] S.-C. Wu, P.-T. Chen, and J.-H. Hsieh, "Spatiotemporal features of electrocardiogram for biometric recognition," *Multidimensional Systems and Signal Processing,* vol. 30, no. 2, pp. 989-1007, 2019.

[13] S.-C. Wu, S.-Y. Wei, C.-S. Chang, A. L. Swindlehurst, and J.-K. Chiu, "A Scalable Open-Set ECG Identification System Based on Compressed CNNs," *IEEE Transactions on Neural Networks and Learning Systems,* 2021.

[14] P.-L. Hong, J.-Y. Hsiao, C.-H. Chung, Y.-M. Feng, and S.-C. Wu, "ECG biometric recognition: template-free approaches based on deep learning," in *2019 41st Annual International Conference of the IEEE Engineering in Medicine and Biology Society (EMBC)*, 2019: IEEE, pp. 2633-2636.

[15] E. Maiorana, P. Campisi, J. Fierrez, J. Ortega-Garcia, and A. Neri, "Cancelable templates for sequence-based biometrics with application to on-line signature recognition," *IEEE Transactions on Systems, Man, and Cybernetics-Part A: Systems and Humans,* vol. 40, no. 3, pp. 525-538, 2010.

[16] M. Savvides, B. V. Kumar, and P. K. Khosla, "Cancelable biometric filters for face recognition," in *Proceedings of the 17th International Conference on Pattern Recognition, 2004. ICPR 2004.*, 2004, vol. 3: IEEE, pp. 922-925.

[17] A. Mahalanobis, B. V. Kumar, and D. Casasent, "Minimum average correlation energy filters," *Applied Optics,* vol. 26, no. 17, pp. 3633-3640, 1987.

[18] A. L. Goldberger *et al.*, "PhysioBank, PhysioToolkit, and PhysioNet: components of a new research resource for complex physiologic signals," *circulation,* vol. 101, no. 23, pp. e215-e220, 2000.